\documentclass[12pt]{amsart}

\newcommand{\ed}{

\end{document}
}

\newcommand{\nin}{\not\in}
\renewcommand{\DH}{\op{DH}}
\newcommand{\DL}{\op{DL}}
\newcommand{\DDH}{\op{DDH}}

\renewcommand{\>}{\rangle}

\newcommand{\bbF}{\mathbb{F}}
\newcommand{\Fstar}[1]{{\bbF_{\!#1}^{*}}}

\newcommand{\be}{\begin{enumerate}}
\newcommand{\ee}{\end{enumerate}}
\newcommand{\bi}{\begin{itemize}}
\newcommand{\ei}{\end{itemize}}
\newcommand{\itm}{\item}
\newcommand{\sm}{\setminus}
\newcommand{\op}[1]{\operatorname{#1}}

\newcommand{\Prob}{\op{Pr}}

\newcommand{\inv}{^{-1}}
\long\def\forget#1\forgotten{}

\newtheorem{thm}{Theorem} 

\newtheorem{prob}[thm]{Problem}
\newtheorem{lem}[thm]{Lemma}
\newtheorem{claim}[thm]{Claim}

\theoremstyle{definition}

\newtheorem{scen}[thm]{Scenario}

\theoremstyle{remark}
\newtheorem{rem}[thm]{Remark}

\title[Fast Diffie Hellman and malicious standards]{Fast generators
for the\\ Diffie-Hellman key agreement protocol\\ and malicious standards}
\author{Boaz Tsaban}
\thanks{Supported by the Koshland Center for Basic Research.}
\address{Boaz Tsaban, Department of Mathematics,
Weizmann Institute of Science, Rehovot 76100, Israel}
\email{boaz.tsaban@weizmann.ac.il}
\urladdr{http://www.cs.biu.ac.il/\~{}tsaban}

\begin{document}
\begin{abstract}
The Diffie-Hellman key agreement protocol is based on taking
large powers of a generator of a prime-order cyclic group.
Some generators allow faster exponentiation.
We show that to a large extent, using the fast generators is
as secure as using a randomly chosen generator. On the
other hand, we show that if there is some case in which fast generators are
less secure, then this could be used by a malicious
authority to generate a standard for the Diffie-Hellman key agreement protocol
which has a hidden trapdoor.
\end{abstract}

\maketitle

\section{Introduction}

The \emph{Diffie-Hellman key agreement protocol} \cite{DH76}
is one of the
most celebrated means for two parties, say Alice and Bob,
to agree on a secret key over an insecure
communication channel.
Alice and Bob make their computations in some previously
fixed cyclic group $G$ with an agreed generator $g$.
The protocol is defined as follows:
\be
\itm Alice chooses a random\footnote{Throughout the paper,
by \emph{random} we mean uniformly random and independent of earlier samples.}
$a\in\{1,\dots,|G|-1\}$,
and sends $g^a$ to Bob.
\itm Bob chooses a random $b\in\{1,\dots,|G|-1\}$,
and sends $g^b$ to Alice.
\ee
The agreed key is $g^{ab}$, which can be computed
both by Alice ($(g^b)^a$) and by Bob ($(g^a)^b$).

Due to the Pohlig-Hellman attack \cite{PH78} (which exploits the
Chinese Remainder Theorem), it is preferred that the order of
the group be prime, which is henceforth assumed.

Consider, for example, the case $g\in\Fstar{q}$ where $q$ is prime.
Let $p$ be the (prime) order of the generated group $G=\<g\>\le\Fstar{q}$.
Computing $g^x$ for $x\in\{1,\dots,p-1\}$ consists of
squaring and multiplying.
If $g=2$, then the multiplication
operation amounts to shifting and taking modular reduction.
For $h\in\Fstar{q}$,
$$2h \bmod q =
\begin{cases}
2h & h<q/2\\
2h-q & q/2\le h
\end{cases}$$
which is computationally negligible in comparison to multiplying by a random $g$.
In standard square-and-multiply implementations this saves about $33\%$ of the computational complexity
of evaluating $g^x$ (in fact, squaring can often be done more efficiently than general multiplication,
so this saves more).
Thus, if $2\in G$,
we may wish to chose it as our generator.
If $2\nin G$, we can use other generators for which similar comments apply (like $3,5$, etc.).

We show that, in the common interpretation,
this can be done with no loss of security.
On the other hand, we show that if there is a conceivable way
to make some generators weaker than random ones,
then this can be used by an authority of standards
to find parameters for the Diffie-Hellman protocol with a trapdoor
allowing the authority to exploit these weaknesses.
In the appendix we give an example of a public-key
cryptosystem based on this phenomenon.

The results also apply to choices of efficient generators
in other groups, e.g., low hamming weight polynomials in
$\Fstar{q^m}$, or low weight elements in hyper-elliptic curves.

\section{A fast generator is almost as secure}\label{secure}

Let $G=\<g\>$ be a cyclic group of prime order $p$.
Let $f\in G$ be any element except the identity.
Then $f$ is a generator of $G$.
In the intended application, $f$ is chosen so that the
computation of $f^x$ is more efficient
(we call $f$ a \emph{fast generator}), or that its usage
is convenient for some other reason.

Fix $h\in G$.
An algorithm $\DH_h$ (depending on $h$) is said to solve
the \emph{Diffie-Hellman Problem (DHP)} for base $h$ if, for each
$x,y\in\{1,\dots,p-1\}$,
$\DH_h(h^x,h^y)=h^{xy}$.

Henceforth, for a number $r\in\{1,\dots,p-1\}$,
$r\inv \bmod p$ denotes the element $s$ of $\{1,\dots,p-1\}$
such that $sr = 1 \pmod p$.

The following theorem is presumably known to specialists,
but we have not been able to find a reference.
The method of proof, however, is standard.

\begin{thm}\label{DHP}
Assume that for some $f\in G\sm\{1\}$, there exists
an algorithm $\DH_f$ to solve the DHP for base $f$,
in running time $T(f)$.
Then for each $g\in G\sm\{1\}$, there is an algorithm $\DH_g$
which solves the DHP for base $g$
in running time $O(T(f)\cdot\log p)$.
\end{thm}
\begin{proof}
Given $g$, there exists a unique $r\in\{1,\dots,p-1\}$ such that
$g=f^r$.
\begin{lem}\label{sqm}
Given $f^r$, we can compute $f^{r\inv \bmod p}$ using at most $2\log p$ queries
to $\DH_f$.
\end{lem}
\begin{proof}
By Fermat's Little Theorem, $r^{p-1}=1 \pmod p$, and therefore
$$r^{p-2}=r\inv \pmod p.$$
We can compute $f^{r\inv}=f^{r^{p-2}}$ using $\DH_f$ in a square-and-multiply
manner: Write $p-2$ in base $2$ as $b_0+b_1\cdot 2+\dots+b_n\cdot 2^n$, $b_n\neq 0$ (then $n\le\log_2p$).
Let
$f_0=f^r$.
For each $i=1,2,\dots,n$ compute
$h_i=\DH_f(f_{i-1},f_{i-1})$,
and let
$f_i=h_i$ if $b_{n-i}=1$, and $f_i=\DH_f(h_i,f_0)$ otherwise.
Then $f_n=f^{r^{p-2}}$.
\end{proof}
Now, assume that we are given $g^x,g^y$ and we wish to find
$g^{xy}$. Recall that $g=f^r$.
Compute $f^{r\inv}$ as in Lemma \ref{sqm},
and proceed with
$$\DH_f(f^{r\inv},g^y)=\DH_f(f^{r\inv},f^{ry})=f^{r\inv ry}=f^y,$$
and
$$\DH_f(g^x,f^y)=\DH_f(f^{rx},f^y)=f^{rxy}=g^{xy}.\qedhere$$
\end{proof}

\begin{rem}[Amplification]\label{probDHP}
Theorem \ref{DHP} generalizes to various other settings.
For example, assume that $\DH_f$ only solves the DHP with probability
$\epsilon$, i.e.,
for each $z\neq xy \pmod p$,
$$\Prob[\DH_f(f^x,f^y)=f^{xy}]\ge \Prob[\DH_f(f^x,f^y)=f^{z}]+\epsilon.$$
Then $\DH_f$ can be transformed to an algorithm which succeeds in probability
arbitrarily close to $1$: Choose random $r,s\in\{1,\dots,p-1\}$,
compute $f^{xr}=(f^x)^r$,  $f^{ys}=(f^s)^y$, and $h=\DH_f(f^{xr},f^{ys})$.
If the output $h$ was correct, then
$$h=f^{xrys}=f^{xyrs}.$$
Let $t=(rs)\inv \pmod p$. Then, in the case of correct output $h$, $h^t=f^{xy}$.
We can repeat this $O(1/\epsilon^2)$ times to get $f^{xy}$
as the most frequent value almost certainly.

Having the algorithm transformed to one which succeeds in probability
very close to $1$, the arguments in the proof of Theorem \ref{DHP} apply.
These assertions apply to all problems mentioned in this paper.
\end{rem}

The closely related
Discrete Logarithm Problem is much easier to deal with:
An algorithm $\DL_h$ is said to solve
the \emph{Discrete Logarithm Problem (DLP)} for base $h$ if, for each
$x\in\{1,\dots,p-1\}$,
$\DL_h(h^x)=x$.

\begin{thm}\label{DLP}
Assume that $f\in G\sm\{1\}$, and there exists
an algorithm $\DL_f$ to solve the DLP
for base $f$, in running time $T(f)$.
Then for each $g\in G\sm\{1\}$, there is an algorithm $\DL_g$
which solves the DLP for base $g$
in running time $O(T(f))$.
\end{thm}
\begin{proof}
Given $g^x$, find $x$ using the following sequence of computations:
$r=\DL_f(g)$, $rx=\DL_f(f^{rx})=\DL_f(g^x)$, $s = r\inv\bmod p$,
and $x=srx$.
\end{proof}

A closely related problem remains open:
An algorithm $\DDH_h$ is said to solve
the \emph{Decisional Diffie-Hellman Problem (DDH)}
for base $h$ \cite{Boneh98} if, for each
$x,y,z\in\{1,\dots,p-1\}$,
$\DDH_h(h^x,\allowbreak h^y,h^z)=1$ if, and only if, $z=xy$.

\begin{prob}
Assume that $f\in G\sm\{1\}$, and there exists
an algorithm $\DDH_f$ to solve the DDH
for base $f$, in running time $T(f)$.
Does there exist, for each $g\in G\sm\{1\}$, an algorithm $\DDH_g$
which solves the DDH for base $g$ in running time polynomial in $T(f)\cdot\log p$?
\end{prob}

\begin{rem}\label{Bleich}
Menezes has pointed out to us that in \cite{Bleich}
it is shown that using $2$ as a generator for certain discrete logarithm based
\emph{signature schemes} is vulnerable to forgeries, whereas
in \cite{PoSte} it is shown that using a random
generator in these schemes is provably secure (this is summarized in \cite{Stern}).
This can be contrasted with the results of the current section,
and motivate the discussions in the remainder of the paper.
\end{rem}

\section{Malicious standards}\label{malicious}

One can still figure out models of security for which it is not
clear that using fast generators is as secure as using a random
generator.
For example, assume that the following holds.

\begin{scen}[Malicious Diffie-Hellman (MDH)]\label{scena}
~\be
\itm There exist $f\in G\sm\{1\}$, a function
$F$, and an efficient algorithm
$\DH_f$ such that for each $x,y\in\{1,\dots,p-1\}$,
$$\DH_f(f^x,f^y)=F(f^{xy}).$$
\itm For a random $g\in G\sm\{1\}$,
$F(g^{xy})$ cannot be efficiently extracted from $g^x$ and $g^y$.
\itm For random $x,y$, $F(f^{xy})$ has enough entropy
to generate a key for symmetric encryption (e.g., $80$ bits).
\ee
\end{scen}

\begin{rem}
While it seems unlikely that MDH could hold, we should note that
the field is full of surprises. For example, in \cite{JouNgu} it
is shown that there are some groups where the Diffie-Hellman
Problem is difficult and the Decisional Diffie-Hellman Problem
(see Section \ref{secure}) is easy.
See Remark \ref{Bleich} for another example.
\end{rem}

If MDH holds,
then $\DH_f$ reveals some information on the agreed key
obtained by the Diffie-Hellman protocol using $f$ as a generator.
In an extreme case, the function $F$ could be the hash function which
Alice and Bob use to derive from $f^{ab}$ a key for symmetric encryption.
However, in general it is not clear
how to use $\DH_f$ to reveal the same information $g^{ab}$ for a random generator $g$.
Of course, there is a random $r\in\{1,\dots,p-1\}$ such that $g=f^r$ and therefore
$$\DH_f(g^a,g^b)=\DH_f(f^{ra},f^{rb})=F(f^{r^2ab})=F(g^{rab}),$$
but $rab$ is a random element of $\{1,\dots,p-1\}$ and independent of $ab$,
so this information is of no use.
Similar assertions hold for the Discrete Logarithm Problem.

Consequently, it might be the case that fast generators are
not as secure as random ones.
While we are unable to prove the impossibility of Scenario \ref{scena},
we can show that if it is possible, then we cannot trust given standards for
the Diffie-Hellman key agreement protocol, unless we know how they
were generated.

Assume that MDH holds.
Then an authority of standards can do the following:
Choose a uniformly random \emph{trapdoor} $t\in\{1,\dots,p-1\}$,
compute $g=f^t$, and suggest $(G,p,g)$ as the standard's
parameters for the Diffie-Hellman key agreement protocol.
As $t$ was uniformly random, $g$ is a uniformly random generator of
$G$, so there is no way to know that it was chosen in a malicious way.
Now, assume that Alice sends Bob $g^a$ and Bob sends Alice $g^b$.
For everyone else but the authority of standards, deducing information
on the agreed key $g^{ab}$ is impossible.

\begin{claim}
For all $a,b\in\{1,\dots,p-1\}$, the authority of standards can compute
$F(g^{ab})$ efficiently.
\end{claim}
\begin{proof}
Using the trapdoor $t$, compute $t\inv \bmod p$, and
$(g^b)^{t\inv}$, which is the same as $f^{tbt\inv} = f^b$.
Now, compute $F(f^{rab})=\DH_f(f^{ra},f^b)$.
But $f^{rab}=g^{ab}$.
\end{proof}

Consequently, the authority of standards can decrypt the messages
sent between Alice and Bob.

\medskip

In the appendix we indicate a possible positive consequence of the MDH.
We believe that many more can be derived from it.
The proof of the impossibility of MDH under mild hypotheses,
or the construction of a system for which MDH holds, are
fascinating challenges.

\begin{rem}
Galbraith has pointed out to us that there exist bit security results
which show that for various natural functions $F$, computing $F(g^{ab})$ from
$g^a$ and $g^b$ is as hard as the Diffie-Hellman Problem.
See, e.g., \cite{ShWi05} and references [1,2] therein.
This is an evidence for the difficulty of establishing MDH.
\end{rem}

\appendix

\section{\\A public-key cryptosystem from the Malicious Diffie-Hellman assumption}

Assume that MDH holds for a group $G$ with prime order $p$ and a generator $f$.
Then we define the following \emph{public-key cryptosystem for celebrities}:
In the intended application, we have some center (a ``celebrity'') sending
messages to many recipients. The purpose is to minimize the communication load
of the center's messages.
\be
\itm $G$ and $p$ are publicly known.
\itm A celebrity, say Bob, chooses a random $r\in\{1,\dots,p-1\}$ and publishes $g=f^r$.
\itm Each one (say, Alice) who wishes to obtain in the future messages from Bob should choose a random
$a\in\{1,\dots,p-1\}$ and publish $g^a$.
\itm When Bob wishes to encrypt a message to Alice, he computes $F(g^{a^2})$ (using $r$ he
can do that, as shown in Section \ref{malicious}) and uses
some known hash function of the result as a key for a block cipher with which he encrypts
the message to Alice.
\itm Alice can compute $g^{a^2}$ and thus decrypt the message.
\itm Users other than Bob who wish to send messages to one another or to Bob can use standard
algorithms like El-Gamal.
\ee
Note that the lengths of Bob's encrypted messages is the same as that of the plain messages.

Our suggested protocol is based on the difficulty of
finding $g^{a^2}$ given $g^a$.
Menezes has pointed out to us that in Section 5.3 of \cite{MauWol} it is shown that
this is as difficult as the Diffie-Hellman Problem:
Indeed, given $g^a$ and $g^b$, compute $g^{a+b}=g^a\cdot g^b$, and then compute
$g^{a^2}$, $g^{b^2}$, and $g^{(a+b)^2}$. Using these, compute
$$g^{2ab}=g^{(a+b)^2}\cdot (g^{a^2})\inv\cdot (g^{b^2})\inv.$$
Finally, compute $g^{ab}=(g^{2ab})^{2\inv\bmod p}$.

\begin{rem}
We can base a protocol with the same properties on classical assumptions:
Bob publishes $g$ \emph{and} $g^b$ (for some random $b$ of his choice), and each other user,
say  Alice, publishes $g^a$ and computes a hash value of $g^{ab}$ to be used as symmetric key
to decipher messages from Bob. Thus, our suggested protocol should only be considered as an indication
of the potential usefulness of MDH, which is not fully understood yet.
\end{rem}

\subsubsection*{Acknowledgments}
We thank Steven Galbraith and Alfred Menezes for their useful comments.

\ed
\begin{thebibliography}{00}

\bibitem{Boneh98}
D.\ Boneh,
\emph{The decision Diffie-Hellman problem},
in: \textbf{Proceedings of the Third Algorithmic Number Theory Symposium},
Lecture Notes in Computer Science \textbf{1423} (1998),
48--63.

\bibitem{Bleich}
D.\ Bleichenbacher,
\emph{Generating ElGamal signatures without knowing the secret key},
Advances in Cryptology -- EUROCRYPT '96,
Lecture Notes in Computer Science \textbf{1070} (1996),
10--18.
(Corrected version: \texttt{ftp.inf.ethz.ch/pub/crypto/publications/Bleich96.ps})

\bibitem{DH76}
W.\ Diffie and M.\ E.\ Hellman,
\emph{New directions in cryptography},
IEEE Transactions on Information Theory \textbf{22} (1976),
644--654.

\bibitem{JouNgu}
A.\ Joux and K.\ Nguyen,
\emph{Separating Decision Diffie-Hellman from Diffie-Hellman in cryptographic groups},
Journal of Cryptology \textbf{16} (2003),
239--247.

\bibitem{MauWol}
U.\ M.\ Maurer and S.\ Wolf,
\emph{The relationship between breaking the Diffie-Hellman protocol and computing discrete logarithms},
SIAM Journal on Computing \textbf{28} (1999),
1689--1721.

\bibitem{PH78}
S.\ Pohlig and M.\ Hellman,
\emph{An improved algorithm for computing logarithms in $GF(p)$ and its cryptographic significance},
IEEE Transactions on Information Theory \textbf{24} (1978),
106--111.

\bibitem{PoSte}
D.\ Pointcheval and J.\ Stern,
\emph{Security Proofs for Signature Schemes},
Advances in Cryptology -- EUROCRYPT '96,
Lecture Notes in Computer Science \textbf{1070} (1996),
387--398.

\bibitem{ShWi05}
I.\ Shparlinski and A.\ Winterhof,
\emph{A hidden number problem in small subgroups},
Mathematics of Computation \textbf{74} (2005),
2073--2080.

\bibitem{Stern}
J.\ Stern,
\emph{The Validation of Cryptographic Algorithms},
Advances in Cryptology -- ASIACRYPT '96,
Lecture Notes in Computer Science \textbf{1163} (1996),
301--310.

\end{thebibliography}
